\documentclass[a4paper,11pt]{article}
\usepackage{amsmath,amsthm,latexsym,amssymb,amsfonts,epsfig}
\usepackage{fontenc,dsfont,layout}
\usepackage{hyperref}
\usepackage{psfrag}
\usepackage[cp1250]{inputenc}
\usepackage{graphics}
\numberwithin{equation}{section}

\DeclareMathAlphabet{\mathpzc}{OT1}{pzc}{m}{it}

\begin{document}
							
\title{ 
\begin{flushright}
\vspace*{-2.0truecm}
{\small CERN-PH-TH/2012-235}
\end{flushright}
\vspace{1truecm}
Dynamical mass scale and approximate scaling symmetry\\ in the Higgs sector}
\author{Zygmunt Lalak\thanks{Zygmunt.Lalak@fuw.edu.pl}}
\date{\it  CERN Physics Department, Theory Division,\\ CH-1211 Geneva 23, Switzerland\\ Institute of Theoretical Physics, Faculty of Physics, University of Warsaw, Ho\.za 69,\\ 00-681 Warsaw, Poland} 
\maketitle

\begin{abstract}
We investigate basic consequences of the assumption that the mass scale of the perturbative sector responsible for the spontaneous symmetry breaking is generated dynamically in a theory with a large UV scale. It is assumed that in addition to an elementary scalar there exists an additional scalar, a modulus, which controls the dynamical hierarchy of scales in the manner similar to that  of supersymmetric gaugino condensation. It is shown that a light degree of freedom appears that couples to the gauge bosons and to charged fermions in a specific way which is different from the couplings of the dilaton of the exact scale invariance. 
\end{abstract}

\section{Introduction} \label{sec:wstep}
Dimensional transmutation in non-abelian gauge sectors is a natural source of mass scales hierarchically smaller than the Planck scale. Such scales can provide relatively small dimensionful coefficients in effective Lagrangians describing extensions of the Standard Modelal at  low energies in the presence of a large UV mass scale. This idea has been used in supersymmetric theories, where hidden sector gaugino condensation sets the scale of the mass splittings in supersymmetry multiplets. The gaugino condensate by itself doesn't break supersymmetry, but it provides a potential for matter superfields, which makes their F-terms non-vanishing, see \cite{Krasnikov:1987jj},\cite{Casas:1990qi}. The well known example  of the dynamical condensation of chiral fermions is the technicolour, where chiral condensate provides masses of the weak gauge bosons without the need for additional elementary scalar fields. 

In this note we  allow for the existence of elementary scalar fields but  follow the idea that mass scales in low energy Lagrangian are generated dynamically. This means that the bosonic part of the Higgs sector of the perturbative Lagrangian is classically scale invariant (see for instance \cite{Coleman:1985}), and that scale invariance is broken via fermion and gluon condensates, as well as due to anomalous contributions. 

\section{Fermionic and gluonic condensates}
For the sake of concretness let's start the discussion with an SU(N) model with quarks in vectorlike pairs. In technicolour models \cite{Weinberg:1975gm,Weinberg:1979bn},\cite{Susskind:1978ms}
one lets  some of the techniquarks $\psi$ to be electroweak doublets in addition to being N-tuplets of SU(N). In this case the condensates break directly the electroweak symmetry. We shall consider in what follows a possibility that techniquarks are electroweak singlets and do not participate directly in the breaking of the $SU(2)_W$. 
There are in principle two types of condensates which may form: chiral condensates $\bar \psi_L \psi_R$ which break spontaneously chiral symmetries and the scale symmetry, and gluon condensates $<F^2>$ which also break the scale symmetry spontaneously.  In addition, scale symmetry is broken explicitly due to anomaly. 
Approximately, the condensate $<F^2>$ forms when $C_2 (F F) \alpha(Q^2=\Lambda^{2}_{D}) = 1$ and the chiral condensate forms when $ C_2 (\bar \psi \psi) \alpha(Q^2 = \Lambda^{2}_\chi) = 1$, where $C_2$ is a quadratic Casimir in respective channels. Obviously, it is natural to expect $\Lambda_D > \Lambda_\chi$ since $C_2$ for adjoint representation is larger than for the fundamental one. More specifically 
\begin{equation}
\frac{\Lambda_D}{\Lambda_\chi} \sim e^{\frac{2 \pi (N-1)}{b \alpha(\Lambda_D) N}}. 
\end{equation}
That is the scale symmetry spontaneous breaking is dominated by the gluon condensate. 
If the scale $\Lambda_D$ is high enough, the coefficient of the anomaly $\beta(\alpha)/\alpha (\Lambda_D)$ is small and the explicit anomalous breaking of the scale symmetry is small wrt spontaneous one. Therefore one expects a scalar pseudo-goldstone boson to appear in the spectrum, which one usualy calls a (techni-) dilaton. 
Deatails are model dependent.  The useful parametrization of the technidilaton is 
via the field $U=F_D e^{\phi/F_D}$ where $\phi$ is the dilaton which transforms as $\phi (x) \rightarrow \phi (\sigma \, x) + F_D \log (\sigma)$ under the scale symmetry, which means that $U$ behaves as a field of canonical dimension $1$. The kinetic term for the dilaton can be written as $1/2 \, (\partial U)^2$. 
One can write down an effective potential for the dilaton below the gluon condensation scale, see \cite{Goldberger:2007zk},\cite{Grinstein:2011dq},\cite{Campbell:2011iw} :    
\begin{equation} \label{an}
V_{an} = \frac{m^2}{16 F^{2}_D} U^4 \left (\log (U/ F_D)^4 -1 \right ),
\end{equation}
This potential has a minimum at $<U> =F_D $ and the mass of $U$ is $m^{2}$. This potential is sufficient to describe the effect of the quantum scale anomaly, and in general is suffcient if the dimension $d^\star =d+ \gamma$ of operators violating scale invariance fulfills the condition $|d^\star -4| \ll 1$. If there are two condensing non-abelian gauge groups, let's label them $1$ and $2$ respectively, then  below the smaller of the two gluon condenation scales there exists a single dilaton, which shifts under the scale symmetry:
\begin{equation}
\phi = \cos(\theta) \phi_1 + \sin(\theta) \phi_2, \,\, {\rm and} \,\, \tan(\theta) = \frac{F_2}{F_1}.
\end{equation}
The dilaton decay constant equals $F= \cos(\theta) \frac{F^{2}_1 + F^{2}_2}{F^{2}_1}$. If one assumes that the effective potential is dominated by the sum of the terms of the form (\ref{an}), both the dilaton and the orthogonal combination of 
$\phi_1$ and $\phi_2$ are stabilized due to the anomaly induced effective potential. 

\subsection{A simple model}
Now we would like to write down a simple Higgs sector where the mass scale is dynamical. Let 
\begin{equation}
{\cal L}(h,\psi) = \frac{1}{M} {\bar \psi} \psi h^2 - \frac{\lambda}{4} h^4,
\end{equation}
where $M$ is a UV mass scale, corresponding to a mass of heavy states that have been integrated out (it could be as high as the Planck scale). The usual flavour problem known from technicolour models puts lower limit on the scale $M$. 
Fermions $\psi$ are SM singlets, so there are no gauge bosons which couple to both visible and hidden fermions. However,  a heavy scalar of mass $M$ could couple to both sectors and induce flavour changing four-fermion operators. To suppress such effects without introducing flavour symmetries one needs to take $M>5 \, {\rm TeV}$, see \cite{Lalak:2010bk},\cite{Calibbi:2012at}. In fact, since there is an elementary scalar to form standard Yukawa couplings in the model, one can raise the scale $M$ even up to to the Planck scale. For the purpose of this note we shall suppress possible gravitational effects. Actually, we are interested in a possibility of maintaining a large hierarchy between the electroweak scale and a high UV scale $M$. 

The perturbative scalar potential is obviously scale invariant, although the scale invariance is broken by the coupling of the scalar $h$ to (techni-) fermions. Below the chiral condensation scale one obtains the effective potential for $h$
\begin{equation} 
V(h)= -\Lambda^{2}_\chi \left (\frac{\Lambda_\chi}{M} \right)^{1+\gamma} h^2 + \frac{\lambda}{4} h^4,
\end{equation}
where $\gamma$ is an anomalous dimension of the operator $O=\bar \psi \psi$. 
Let's assume that the fundamental theory above condensation scale is scale invariant. Therefore one can make the Higgs Lagrangian scale invariant multiplying the effective mass term by a suitable power of $U$ 
\begin{equation}
V(h)= -\Lambda^{2}_\chi \left (\frac{\Lambda_\chi}{M} \right)^{1+\gamma} e^{- (1+\gamma) \phi /F_D} h^2 + \frac{\lambda}{4} h^4,
\end{equation}
plus a scale invariant kinetic term for $\phi$. 
The effective Lagrangian which includes couplings of the Higgs to the longitudinal parts of the gauge bosons looks as follows
\begin{equation}
{\cal L} = \frac{1}{4} D_\mu \Sigma D^\mu \Sigma e^{2 \phi/F_D} h^2 + \frac{1}{2} (\partial U)^2 - V(h, \phi) - V_{an} (\phi).
\end{equation}
Minimizing with respect to $h$ one obtains
\begin{equation}
<h^2> = \frac{2 \Lambda^{2}_\chi}{\lambda} \left ( \frac{\Lambda_\chi}{M} \right )^{1+\gamma} e^{-(1+\gamma) \phi/F_D},
\end{equation}
and, upon substitution the above expression into the couplings of the Higgs to gauge bosons, one finds the coupling of the dilaton to the SM gauge bosons:
\begin{equation}
{\cal L} = \frac{1}{4} D_\mu \Sigma D^\mu \Sigma v^2 \left ((1-\gamma) \bar U /F_D + \frac{1}{2} (\gamma - 1) \gamma {\bar U}^2 / F^{2}_D + ... \right ), 
\end{equation}
where $v^2 = \frac{ 2 \Lambda^{3+ \gamma}_\chi }{\lambda M^{1+\gamma} } $ and $\bar U = U - F_D$. One should note that this expression differs from the one usually quoted for the dilaton
$v^2 (2 \bar U /F_D + {\bar U}^2 / F^{2}_D )$, which holds also for the coupling of the SM Higgs field to the gauge bosons,  even in the case of negligible $\gamma$. 
Now the couplings to fermions arise via the anomalous coupling to the trace of the energy-momentum tensor, $\delta L_U = \bar U / F_D \, T^{\mu}_{\mu}$, which includes further coupling to the standard model particles. 
Properties of particles with such couplings have been studied in the light of available data in the papers \cite{Azatov:2012bz},\cite{Ellis:2012hz},\cite{Low:2012rj}. 

\section{Models with a modulus}
However,  it is reasonable to consider more complicated situations, beyond the simplest model described above. In string inspired models there exist scalar moduli fields, which are flat directions of the scalar potential and whose expectation values set the magnitude of various couplings, including the gauge couplings. 
Let us assume the following scalar field dependence of the gauge kinetic term 
\begin{equation}
{\cal L}_{kin} =\frac{1}{4} (\Delta_i + \frac{s}{8 \pi^2}) F^{2}_{i} , 
\end{equation}
where $i$ labels factors in the gauge group. In principle one should write $s_i$ - there could be an independent dilaton/modulus for each  factor. For the dilaton, these  couplings would arise as anomalous couplings to the gauge bosons. In what follows we assume that there is a single modulus entering each gauge coupling. 
The definition above corresponds to the normalization $<s>= \frac{8 \pi^2}{g^2 (\Lambda_{UV})}$. 
With this choice the dynamical gauge coupling the 1-loop RGE invariant condensations scale in a strongly interacting gauge sector becomes
\begin{equation}
\Lambda_{1-loop}= \Lambda_{UV} e^{-s/b},
\end{equation}
where the threshold correction $\Delta$ has been swallowed by the redefinition of $\Lambda_{UV}$. This means that the UV cutoff scale could be slightly different for different condensing gauge groups. In what follows we shall identify the UV cut-off with $M$ - the mass of heavy particles integrated out to produce operators $h^2 \psi^2, \; (\psi^2)^p$, hence $\Lambda_{UV} = M$. 
The expectation value of $s$ has the interpretation of the initial condition for the running at the UV scale $M$, which we shall try to select dynamically with the help of the low energy effective Lagrangian. Alternatively, one could trade the variable $s$ for one of the dynamical scales, say $\Lambda = \Lambda_1$. In this case one can express the condensation scales of all the groups via the scale $\Lambda$ as follows:
\begin{equation}
\Lambda_i = M \left ( \frac{\Lambda}{M} \right )^{b_i /b_1},
\end{equation}
where $i$ labels the gauge factors. 
Hence, variation of the effective potential with respect to $s$ is equivalen to the variation with respect to the dynamical scale $\Lambda$. Still another point of view comes from the fact that 
\begin{equation}
\frac{1}{4} (\Delta_i + \frac{s}{8 \pi^2}) F^{2}_{i} = \frac{1}{4} (\Delta_i + \frac{b_i}{8 \pi^2} \log \left ( \frac{M}{\Lambda_i} \right ) ) F^{2}_{i} . 
\end{equation}
This means that stability with respect to variations of $s$ corresponds to stability with respect to variations of the UV cut-off scale $M$. 

Assuming the relevant condensates are of the form $F^2$ and $(\bar \psi \psi)^p$ the most general effective potential one expects below all condensaton scales takes the form\footnote{For simplicity we shall omit the label $\chi$ on $\Lambda_\chi$.}
\begin{equation} \label{general_pot}
V(h,s)= -\Lambda^{2} \left ( \frac{\Lambda}{M} \right )^{1+\gamma} h^2 + \frac{\lambda}{4} h^4 + 
\sum_i \epsilon_i \Lambda_{\, i}^4 \left ( \frac{\Lambda_{\, i} }{M} \right )^{\gamma_i} + \sum_{i=1, \, p=2}^{\infty} \epsilon_{i,p} \Lambda_{\, i}^4 
\left ( \frac{\Lambda_{\, i} }{M} \right )^{3 p + \gamma_{i, \, p} -4}, 
\end{equation}
with $\epsilon_x$ denoting real numbers of order 1 and  of various signs, and $\gamma$s being anomalous dimensions of respective operators. 
In general, one expects only smallest powers of the scales to contribute to vacuum stabilization. Moreover, the smallest possible number of independent condensing sectors should be emlpoyed. It is rather obvious, that one cannot achieve much with a single condensing sector: one would need a cancellation between terms of different order in $\Lambda$, which under assumption of order 1 coefficients would imply $M\sim\Lambda$, hence no small parameter and no hierarchy of scales in the model. Therefore, we shall restrict ourselves to 2 independent condensates. In fact, as will become clear soon, interesting solutions are determined by pairs of condensing sectors with very close condensation scales, hence the restriction to two condensates is not very restrictive. 
To start the discussion on a familiar ground, let's start with the simple supersymmetric case. 

\section{Low energy supersymmetric Higgs sector with a dynamical scale}

Let us have a look at a supersymmetric extension of the Higgs sector. One knows, that the mass term mixing the up- and down- type Higgs superfields is allowed and its magnitude should be of the order of the electroweak scale. Let's assume that that the $\mu$-term is generated dynamically. The simplest superpotential that can achieve the goal is 
\begin{equation}
W= -M^3 e^{- \frac{3 s }{b_1 M} } + B M^3 e^{- \frac{3 s}{b_2 M}} + M e^{- \frac{3 s }{b_1 M} } H_1 H_2.
\end{equation}
Again, let's assume for simplicity the canonical kinetic term for the modulus $s$. It is well know that a noncanonical kinetic term would introduce additional problems, but let us concentrate on the hierarchy generation itself.  
We are going to assume that the sole role of the condensate is to create a hierachically small mass scale and assume that soft terms are generated in a different sector of the model. The scalar potential in the Higgs sector for the neutral components of the fields looks as follows: 
\begin{equation}
V_h = M^2 L^{6}_1 (h^{2}_1 + h^{2}_2) + 9 \left ( (h_1 h_2 -M^2) L^{3}_1 /b_1 + M^2 B L^{3}_2 / b_2 \right )^2  + m^{2}_1 h^{2}_1 + m^{2}_2 h^{2}_2 + b h_1 h_2 + 
\frac{g^{2}_2}{8} \left ( h^{2}_1 - h^{2}_2 \right )^2, 
\end{equation}
where $L_1= e^{- \frac{ s }{b_1 M} }$.
The above includes soft terms, $m^{2}_1, \, m^{2}_2, \, b$, but it is assumed that these soft terms are $s$ independent, since the dynamics of supersymmetry breaking doesn't have to depend on $s$. However, the $g_2$ can be proportional to  $1/s$. Then the $D$-term  tends to drive the modulus towards infinity, unless the dynamical scale  $\mu = M L^{3}_1$ turns out larger than $<D>$. In what follows we shall assume that is the case. In this regime the scalar potential has a local minimum at 
\begin{equation}
<s> = \frac{M b_1 b_2}{b_1 - b_2} \log \left ( \frac{b_1 B}{b_2}  \right ). 
\end{equation}
To obtain a large hierarchy with respect to $M$ one needs to generate a large value of $s$, which can be naturally achieved via a discrete choice of the beta function coefficients $b_i$, while keeping other coefficients of order one. 
The light eigenvalue of the mass matrix turns out to be mostly along the direction of $s$:
\begin{equation}
m^{2}_s = 162 M^2 \frac{(b_1 - b_2)^2}{b^{4}_1 b^{2}_2 } L^{6}_1 = 162 \mu^2 \frac{(b_1 - b_2)^2}{b^{4}_1 b^{2}_2 } . 
\end{equation}
Thus it is suppressed wrt to the tree-level Higgs mass parameter $\mu$ but not by additional powers of $L$ but by a small combination of $b$s. It is obvious that to achieve a large value of $s$ one needs to take $b_1 \sim b_2$, which makes the combination $\frac{(b_1 - b_2)^2}{b^{4}_1 b^{2}_2 }$ naturally small, however the suppression is rather power-like than exponential. The point is that in the limit $b_1=b_2$ the superpotential has a symmetry $s\rightarrow s + \delta$ which holds up to the overall rescaling of the Lagrangian (hence it is a symmetry of equations of motion). In that limit the lightest eigenvalue is exactly zero, but of course one never actually takes this limit and the shift symmetry is broken. However, the scaling symmetry under which the Higgs fields scale is broken badly by the purely nonperturbatibe terms.  As a byproduct one also obtains a superpotential correction to the Higgs' quartic couplings $\delta V_4 = 9 L^{6}_1 h^{2}_1 h^{2}_2$ which is subleading wrt $g^{2}_2$. Thus below the scale of the electroweak breaking one finds othe usual structure known from the minimal supersymmetric extensions of the Standard Model plus a  light scalar which couples to $Z$ via 
\begin{equation}
\delta v^2 = ... + \frac{96}{g^{2}_1 + g^{2}_2} \frac{1}{b_1} \frac{\mu^2}{M} \delta s . 
\end{equation}
 This scalar can also couple to photons if the gauge kinetic terms contain the linear coupling to $s$. 

\section{Nonsupersymmetric models with dynamical scales}

Let us simplify the general scalar potential (\ref{general_pot}) and retain only two condensing gauge groups and operators of mass dimension 6, thus neglecting gluon condensates. Another point of view is that we have assumed that the dilaton has been stabilized by the gluonic condensation and omit its effective potential. Furthermore, let's take the case of small anomalous dimensions. 
In this case the general form of an effective potential becomes 
\begin{equation} \label{pots}
V(h)= - \left ( \frac{\Lambda^{3}_{1}}{M}  \right ) h^2 + \frac{\lambda}{4} h^4 
- \frac{\Lambda^{6}_{1}}{M^2} + B \frac{\Lambda^{6}_{2}}{M^2},
\end{equation}
where labels $1,2$ correspond to two different condensing groups, and only one of them is allowed to couple to $h^2$, $B$ is an order $1$ coefficient, 
$M=\Lambda_{UV}$ and the phases of the two chiral condensates have been chosen in a convenient way.  
Lets us assume that the chiral condensation scale equals $\Lambda_{1-loop}$: 
\begin{equation} \label{potss}
\Lambda_{}=M e^{-s/(M b)}. 
\end{equation}
The Higgs potential has an approximate scale symmetry broken by $\Lambda_{\, 2}^6$ term where the modulus $s$ shifts under the symmetry transformation, i.e. $h(x)\rightarrow \sigma h(\sigma x)$ and $s(x) \rightarrow s(\sigma x) - 2/3 \, 
M b \log (\sigma)$.  
The scale invariant kinetic term for $s$ would be 
$\frac{1}{2} \partial U \partial U, $
with $U=M e^{-\frac{3 s}{2 M b}}$. However, the modulus in question is not a dilaton, so taking a non-invariant canonical kinetic term for $s$ seems to be more appropriate. 
Substituting \ref{potss} into \ref{pots} one finds a nontrivial potential for both $s$ and $h$. The question is whether one can generate this way a suitable spectrum of masses, stable under perturbative corrections. The point is that $s$ and $h$ would mix, and both mass eigenstates will couple to fermions, and gauge bosons in a higgs-like manner and additionaly via "anomalous" couplings $\sim \, s F^2$ if $s$ is an universal modulus which couples to visible gauge fields as well.   
 
 The model can be solved exactly. One finds a minimum at 
 \begin{equation}
<h^2> = \frac{2 M^2 L^{3}_1}{\lambda}, \,\, <s> = M \frac{b_1  b_2}{6 (b_1 - b_2)} \log \left ( \frac{B b_1}{b_2} \frac{1}{1+ 1/\lambda} \right ). 
\end{equation}
In the above one needs to take $b_1 > b_2$ and $B > \lambda / (1+ \lambda)$ and $L_{1,2} = e^{-s/(M b_{1,2})}$. Obviously, to obtain large hierarchy one needs $b_{1,2}$ close to each other, as in the supersymmetric racetrack models. This means that in models with hierarchy $L_1 \sim L_2$. One can find the mass eigenvalues which are
\begin{equation}
m^{2}_2 = L^{3}_1 m^{2}_1, \,\, m^{2}_1 = 4 M^2 L^{3}_1 . 
\end{equation}
Now, if we want to give the electroweak gauge bosons a mass through the vev of $h$, we need 
\begin{equation}
<h^2> = 2 M^2 L^{3} / \lambda = (246 \,{\rm GeV} )^2.
\end{equation}
Obviously, it is possible to identify the eigenstate of the larger mass $m_1$ with the Higgs particle seen at $125 \, {\rm GeV}$ while keeping  $\lambda$ small. 
This implies 
\begin{equation}
\lambda = 0.13 \,\, {\rm and} \,\, L_1 = 10^{-11} . 
\end{equation}
When one identifies the ultraviolet scale $M$ with the Planck scale, one finds $m_2 = 1.25 \times 10^{-5.5} \, {\rm eV}$. The identification of $m_2$ with the Higgs mass of $125 \, {\rm GeV}$ would require a large value of $\lambda$, since the ratio of the two masses would need to obey the constraint
$m^{2}_2/m^{2}_1 = 2 L^{3}_1 \lambda = (125/246)^2$, which requires a large value of $\lambda$ to cancel the hierarchy factor $L^{3}_1$.  
On the other hand, one could lower the scale $M$ and make the hierarchy smaller, to obtain a smaller, more natural,  $\lambda$.
Thus the model has a clear prediction: it can account for the observed Higgs mass and for the correct electroweak breaking, but the price to pay is a very light scalar in the spectrum which couples "anomalously" to visible gauge bosons. 

\subsection{Perturbative corrections to the mass hierarchy} 
Since there is a very light state in the spectrum, one needs to consider 1-loop corrections to the scalar potential. We take into account Higgs and top quark loops. 
Their contribution is as follows
\begin{equation}
V_{1-l} (h,s) = \frac{1}{64 \pi^2} \left ( (3 \lambda h^2 - 2 M^2 L^{3}_1 )^2 \log \left ( (3 \lambda h^2 - 2 M^2 L^{3}_1 )/Q^2 \right ) - 4 y^{4}_t h^4 \log \left ( y^{2}_t h^2 / Q^2 \right ) \right ), 
\end{equation}
where $y_t$ is the top quark Yukawa coupling and $Q$ is the renormalization scale. Obviously this correction will not affect significantly the mass of $h$ (that is $m_1$) but it could introduce corrections to 
the light mass $m_2$. To see the scale of these corrections it is sufficient to find the derivatives of $V_{1-l}$ wrt $s$. One finds
\begin{equation}
\frac{\partial V_{1-l} }{\partial s} = \frac{3}{8 \pi^2 b_1} M^3 L_{1}^6 \left ( 1 + 2 \log \left (4 L^{3}_1 M^2/Q^2 \right )  \right ). 
\end{equation}
This should be compared to the derivative of the tree level potential
\begin{equation}
\frac{\partial V_{0} }{\partial s} = \frac{6 M^3 L^{6}_1}{b_1} (1+ 1/\lambda) - \frac{6 B M^3 L^{6}_2}{b_2} .
\end{equation}
It is clear that we can minimize the 1-loop correction by choosing the the perturbative renormalization scale $Q^2 \sim M^2 L^{3}_1\sim \lambda v^2$. In this case only numerical coefficients will be changed with respect to the tree-level results. In particular, the 1-loop correction to $m^{2}_2$ is $\delta_{1-l} m^{2}_2 = \frac{9}{16 \pi^2} \frac{ M^2 L^{6}_1 }{b^{2}_1}$. Corrections coming from the the SM gauge bosons are propotrional to $\log(\frac{g^2}{\lambda^2})$ which is not large in the present case, hence they do not modify the overal picture.  

Hence, in this case one can built and extended Higgs sector with a large hierarchy between scalar mass eigenstates, which allows to make the lighter state much smaller than $100$ GeV. This hierarchy can be understood as the result of an approximate scale invariance of the effective potential. 

\subsection{Operators of mass dimension 4}
Let us consider for comparison with the previous case the following scalar potential
\begin{equation} \label{pots4}
V(h)= - \left ( \frac{\Lambda^{3}_{1}}{M}  \right ) h^2 + \frac{\lambda}{4} h^4 
- \Lambda^{4}_{1} + B \Lambda^{4}_{2}.
\end{equation}
Here one easily finds a minimum at $<h^2> = 2 M^2 L^{3}_1 /\lambda$ and $s$ given by $\Lambda^{4}_1 / b_1 = B \Lambda^{4}_2 / b_2$. 
The scalar mass eigenstates are 
\begin{equation}
m^{2}_1 = 4 L^{3}_1 M^2, \,\,\, m^{2}_2 = 16 \frac{b_1 - b_2}{b^{2}_1 b_2} L^{4}_1 M^2 .
\end{equation}
Here the hierarchy of mass eigenstates is rather mild, $m_2 / m_1 \sim \sqrt{L_1}$.  

\subsection{Higher dimension operators}

Let' consider a more general potential 
\begin{equation}
V(s,h)= - M^2  h^2 L^{3 + \gamma}_1 + \frac{\lambda}{4} h^4 
- M^4 ( \Lambda^{p_1+ \gamma_1}_{1} - B \Lambda^{p_2 + \gamma_2}_{2}),
\end{equation}
with $p_i \geq 6$. In the equation $V_s=0$ the second term above can be omitted to a good approximation in comparison to the first one. 
The creation of the minimum with the first two terms (after changing the sign in the second one) doesn't work, since it would imply $M\sim\Lambda_1$. 
However, one finds a minimum at 
\begin{equation}
<h^2 > = \frac{2 M^2 L^{3+\gamma}_1}{\lambda}, \; \frac{s}{M} = \frac{b_1 b_2}{(p_2 + \gamma_2) b_1 - (6 + 2 \gamma ) b_2} 
\log \left ( \frac{B(p_2 + \gamma_2) b_1 \lambda}{(6 + 2 \gamma) b_2} \right ). 
\end{equation}
Now, a trivial option is to build a hierarchy with the help of a large logarithm, which amounts to a large tuning. 
The alternative option is to rely on the smallnes of a  denominator in the solution for $s$, which means 
\begin{equation}
p_2 + \gamma_2 \approx (6 + 2 \gamma) \frac{b_2}{b_1}. 
\end{equation}
The mass eigenstates of the mass matrix are 
\begin{equation}
m^{2}_1 =4 M^2 L^{3+ \gamma}_1, \; m^{2}_2 = \frac{(6+2 \gamma) }{\lambda b^{2}_1 b_2} M^2 L^{6 + 2 \gamma}_1 ( (p_2+\gamma_2) b_1 - (6 + 2 \gamma) b_2). 
\end{equation}
One can see, that here the $m^{2}_2 \sim m^{2}_1 L^{3+ \gamma}_1( (p_2+\gamma_2) b_1 - (6 + 2 \gamma) b_2) / (b_1 b_2) $. The term in paranthesis is small if the $L_1$ is small, 
hence the hierarchy of the mass eigenvalues may be large. This is unaffected by 1-loop corrections if the renormalization scale is properly chosen: $Q\approx 4 L^{3+\gamma}_1 M^2$. Up to corrections coming from the anomalous dimensions once one arranges for a large hierarchy between $M$ and $\Lambda_\chi$, one finds an approximate scale symmetry of the potential broken only by the departure of $p_2/b_2$ from $6/b_1$.  
Again, if one wishes to identify one of the eigenstates with the $125$ GeV Higgs particle, it should be the heavier one. In the opposite case 
\begin{equation}
m^{2}_2 = v^2 L^{3 + \gamma}_1 \frac{ (3 + \gamma) ( (p_2+\gamma_2) b_1 - (6 + 2 \gamma) b_2)}{b^{2}_1 b_2}
\end{equation}
and one needs to abandon the hierarchy between $M$ and $v$. 

\subsection{Coupling of the pseudogoldstone scalar to gauge bosons and fermions}
The leading dependence of the couplings of the modulus $s$ to the gauge bosons and fermions can be read from the expressions for $<h^2>$. Using the popular low energy parametrization 
\begin{equation}
{\cal L} = \frac{1}{4} D_\mu \Sigma D^\mu \Sigma v^2 \left (1+ 2 a \frac{s}{v} + b (\frac{s}{v})^2 + ...\right ) - \sum_q m_q \bar q q \left (1 + c_q \frac{s}{v}  \right ), 
\end{equation}
one finds $a=c$, $b= 2 a^2$ and 
\begin{equation}
a = - \frac{(3 + \gamma) v}{b_1 M}. 
\end{equation}
In fact, there is a mixing between the eigenstates of the mass matrix, but as long as $L_1 \ll 1$, the mixing angle is very small. 
\section{Summary}
In this paper we have explored basic features of the assumption that the mass scale of the perturbative sector responsible for the spontaneous symmetry breaking is generated dynamically out of a large UV mass scale. We have assumed that in addition to an elementary scalar there exists an additional scalar, a modulus, which controls the dynamical hierarchy of scales, like in the case of supersymmetric gaugino condensation. This modulus can be indentified with a dynamical degree of freedom setting the high energy cut-off scale M. We have found an approximate scale symmetry of the effective Higgs sector which is related to the dynamical hierarchy of scales. Generically, there appears a light degree of freedom which couples to gauge bosons and to light fermions. The couplings of this scalar are different from those of the dilaton of the exact scale invariance. Although no specific microscopic realization has been given, a number of general features of such a scenario has been enumerated. 
The basic difference with respect to the scenarios based on exact scale invariance is that the scaling-like symmetry is from the start only approximate and restricted to the effective Higgs sector, at low energies it is broken by dynamically generated mass scales.

\begin{center}
{\bf Acknowledgements}
\end{center}
Author thanks G. G. Ross for very helpful discussions and S. Pokorski and D. M. Ghilencea for useful comments. \\
This work was partially supported by Polish Ministry for Science and Education under grant N N202 091839, by National Science Centre under research grant DEC-2011/01/M/ST2/02466 and by the European Community's Seventh Framework Programme under grant agreement PITN-GA-2009-237920 (2009-2013).


\begin{thebibliography}{99}
\bibitem{Krasnikov:1987jj}
  N.~V.~Krasnikov,
  Phys.\ Lett.\ B {\bf 193} (1987) 37.
  
\bibitem{Casas:1990qi}
  J.~A.~Casas, Z.~Lalak, C.~Munoz and G.~G.~Ross,
  Nucl.\ Phys.\ B {\bf 347} (1990) 243.

\bibitem{Coleman:1985}  
S.~Coleman, "Dilatations", in "Aspects of symmetry" (Cambridge Univ. Press, Cambridge, 1985) p. 67.

\bibitem{Weinberg:1975gm}
  S.~Weinberg,
  Phys.\ Rev.\ D {\bf 13} (1976) 974.
  
\bibitem{Weinberg:1979bn}
  S.~Weinberg,
  Phys.\ Rev.\ D {\bf 19} (1979) 1277.
  
\bibitem{Susskind:1978ms}
  L.~Susskind,
  Phys.\ Rev.\ D {\bf 20} (1979) 2619.
  
\bibitem{Goldberger:2007zk}
  W.~D.~Goldberger, B.~Grinstein and W.~Skiba,
  Phys.\ Rev.\ Lett.\  {\bf 100} (2008) 111802
  [arXiv:0708.1463 [hep-ph]].

\bibitem{Grinstein:2011dq}
  B.~Grinstein and P.~Uttayarat,
  JHEP {\bf 1107} (2011) 038
  [arXiv:1105.2370 [hep-ph]].

\bibitem{Campbell:2011iw}
  B.~A.~Campbell, J.~Ellis and K.~A.~Olive,
  JHEP {\bf 1203} (2012) 026
  [arXiv:1111.4495 [hep-ph]].

\bibitem{Lalak:2010bk}
  Z.~Lalak, S.~Pokorski and G.~G.~Ross,
  JHEP {\bf 1008} (2010) 129
  [arXiv:1006.2375 [hep-ph]].
  
\bibitem{Calibbi:2012at}
  L.~Calibbi, Z.~Lalak, S.~Pokorski and R.~Ziegler,
  JHEP {\bf 1207} (2012) 004
  [arXiv:1204.1275 [hep-ph]].

\bibitem{Azatov:2012bz}
  A.~Azatov, R.~Contino and J.~Galloway,
  JHEP {\bf 1204} (2012) 127
  [arXiv:1202.3415 [hep-ph]].

\bibitem{Ellis:2012hz}
  J.~Ellis and T.~You,
  arXiv:1207.1693 [hep-ph].
  
\bibitem{Low:2012rj}
  I.~Low, J.~Lykken and G.~Shaughnessy,
  arXiv:1207.1093 [hep-ph].
     
\end{thebibliography}
\end{document}